\documentclass[conference]{IEEEtran}
\IEEEoverridecommandlockouts
\usepackage{cite}
\usepackage{amsmath,amssymb,amsfonts}
\usepackage{algorithmic}
\usepackage{graphicx}
\usepackage{textcomp}
\usepackage{xcolor}
\usepackage{hyperref}
\usepackage{xurl}
\usepackage{subcaption}
\usepackage{caption}

\def\BibTeX{{\rm B\kern-.05em{\sc i\kern-.025em b}\kern-.08em
    T\kern-.1667em\lower.7ex\hbox{E}\kern-.125emX}}

\IEEEoverridecommandlockouts
\IEEEpubid{\makebox[\columnwidth]{ 
979-8-3503-6431-6/24/\$31.00 \copyright2024 IEEE
\hfill}
\hspace{\columnsep}\makebox[\columnwidth]{ }} 
\begin{document}

\title{Agent-Based Decentralized Energy Management of EV Charging Station with Solar Photovoltaics via Multi-Agent Reinforcement Learning \\
\thanks{*Corresponding author: Hao Wang (hao.wang2@monash.edu).}
\thanks{This work was supported in part by the Australian Research Council (ARC) Discovery Early Career Researcher Award (DECRA) under Grant DE230100046.}
}

\author{
\IEEEauthorblockN{Jiarong Fan}
\IEEEauthorblockA{\textit{Department of Data Science and AI} \\
\textit{Faculty of IT, Monash University}\\
Melbourne, VIC 3800, Australia \\
jiarong.fan@monash.edu}
\and
\IEEEauthorblockN{Chenghao Huang}
\IEEEauthorblockA{\textit{Department of Data Science and AI} \\
\textit{Faculty of IT, Monash University}\\
Melbourne, VIC 3800, Australia \\
chenghao.huang@monash.edu}
\and
\IEEEauthorblockN{Hao Wang*}
\IEEEauthorblockA{\textit{Department of Data Science and AI} \\
\textit{Faculty of IT and Monash Energy Institute} \\
\textit{Monash University}\\
Melbourne, VIC 3800, Australia \\
hao.wang2@monash.edu}
}

\maketitle

\begin{abstract}
In the pursuit of energy net zero within smart cities, transportation electrification plays a pivotal role.  The adoption of Electric Vehicles (EVs) keeps increasing, making energy management of EV charging stations critically important. While previous studies have managed to reduce energy cost of EV charging while maintaining grid stability, they often overlook the robustness of EV charging management against uncertainties of various forms, such as varying charging behaviors and possible faults in faults in some chargers. To address the gap, a novel Multi-Agent Reinforcement Learning (MARL) approach is proposed treating each charger to be an agent and coordinate all the agents in the EV charging station with solar photovoltaics in a more realistic scenario, where system faults may occur. A Long Short-Term Memory (LSTM) network is incorporated in the MARL algorithm to extract temporal features from time-series. Additionally, a dense reward mechanism is designed for training the agents in the MARL algorithm to improve EV charging experience. Through validation on a real-world dataset, we show that our approach is robust against system uncertainties and faults and also effective in minimizing EV charging costs and maximizing charging service satisfaction.

\end{abstract}


\section{introduction}
Smart cities, the future orientation of urban development, emphasize sustainability, efficiency, and enhanced quality of life through the integration of advanced technologies like Information and Communication Technology (ICT), data-driven decision making, and Artificial Intelligence (AI) \cite{silva2018towards,o2019smart}. A key component of this vision is the pursuit of energy net zero to meet smart cities' energy consumption primarily from renewable energy \cite{kylili2015european}. As part of the effort of achieving net zero, transportation electrification plays a vital role, in particular the promotion of widespread adoption of Electric Vehicles (EVs) \cite{iea2020global}.

EV adoption continues rising and is expected to reach around 17 million by the end of 2024, where the market shares of EVs in China, Europe, and the United States may reach up to 45\%, 25\%, and over 11\%, respectively, accounting for more than 20\% cars sold worldwide \cite{IEA2024}. Therefore, the need for reliable and sustainable charging infrastructure, along with EV charging management, becomes more critical than ever before. To ensure that the integration of EVs can reduce Green House Gas (GHG) emissions and enhance the stability of energy systems, effective energy management of EV charging stations, such as coordinating Renewable Energy Sources (RES) and EV charging, is becoming essential to achieving energy net zero for smart cities \cite{bonsu2020towards,wu2021residential,li2024investigating}.

Energy management of EV charging stations initially focused on meeting charging demands for essential operations \cite{bai2010optimum}, which lacked a comprehensive view of the energy system with other resources. Recent studies have expanded the scope to consider enhanced operations, including optimizing EV charging to shave peak load, integrating RES in the charging station, and jointly controlling other flexible resources like battery energy storage \cite{yan2018optimized}. As time-varying resources are considered, energy management tasks have evolved from static to dynamic ones, with the ability to respond to dynamic supply and demand conditions \cite{zheng2018online}. In terms of methodologies, there is a growing application of smart algorithms for EV charging station management using optimization and AI techniques. These methods, e.g., in \cite{jiang2023network,fan2023marl}, leveraging prediction and data-driven approaches, are able to take into account multiple factors, such as charging needs, EV mobility, and energy price fluctuations.

Many smart algorithms for EV charging station management followed the centralized approach. More specifically, centralized approaches process all the information needed to solve the problem, such as charging demands, available energy resources, and electricity prices, and then make decisions for all EVs and energy resources~\cite{jiang2023network,cao2021smart,he2012optimal,zhang2021distributed}. Despite the wide use of centralized approaches, they may face increasing computational challenges as the dimension of the problem grows, making them difficult to scale up.
In contrast, decentralized approaches, through decentralized decision-making, can be more effective in addressing EV charging station energy management with a large group of EVs and other resources.
For example, \cite{liao2021decentralized} proposed a decentralized algorithm using rule-based methods to operate an EV charging station. The flexibility of EV batteries is harnessed through Grid-to-Vehicle (G2V) and Vehicle-to-Grid (V2G), aimed at maximizing operational profits of the charging station. But rule-based methods may lack adaptability, potentially leading to sub-optimal decisions in complex environments due to their reliance on predefined logic or rules.
% optimization
More systematic methods were developed using decentralized optimization and control theories, e.g., in \cite{paudel2022decentralized} and \cite{torreglosa2016decentralized}. More specifically, \cite{paudel2022decentralized} introduced a cooperative strategy based on optimization for EV charging, following the consensus concept and assuming peer-to-peer information sharing among EVs.
% predictive control
Following a similar approach, \cite{torreglosa2016decentralized} employed model predictive control to manage energy supply from the photovoltaic (PV) system, battery, and grid for EV charging, executing this in a decentralized fashion. 
However, these methods either rely on extensive communications between EVs to find a solution or require accurate forecasts to make quality decisions, which may not be feasible in real-world environments.

Agent-based methods have demonstrated great promise in large-scale decentralized decision-making, where agents use local information and interact with other agents to solve complex problems \cite{karfopoulos2012multi}.
% RL for agent
Among various agent-based decision-making strategies, Reinforcement Learning (RL) has gained popularity, as it enables agents to interact with an environment to learn their policies or strategies that are adaptive and robust in handling complex problems with uncertainties \cite{sutton2018reinforcement}.
% RL + EV
In recent years, many studies have focused on RL for EV charging station management \cite{cao2021smart,abdullah2021reinforcement,liu2019smart,wang2019reinforcement,ding2020optimal}.
% single AGENT
But single-agent methods adhere to the centralized structure for managing EV charging stations, thus inheriting the limitations of centralized approaches as discussed above.
Multi-Agent Reinforcement Learning (MARL) instead becomes a suitable method for managing EV charging stations, with great potential to provide effective coordination and improve scalability \cite{fan2023marl,zhang2021intelligent,zhang2022multistep,fu2023electric,li2024decentralized}. Related works primarily focused on privacy preservation of EV charging, as the most mentioned motivation of using MARL, but often overlooked practical challenges that MARL can potentially address. For example, real-world deployment of EV charging stations frequently encounters faults or failures in EV chargers. Both centralized approaches and coordination-based approaches could fail in such scenarios, thus compromising the reliability of the EV charging station and leading to economic losses. In addition, existing studies often overlooked user experiences in EV charging management. From EV users' perspective \cite{UDDIN2018342}, whether charging management can take care of battery degradation and provide satisfactory charging services receives considerable attention in addition to charging costs.

Our research aims to fill in existing gaps by crafting an agent-based, decentralized energy management strategy for EV charging stations, tailored to more realistic scenarios. For example, faults could happen to some chargers, and the charging station is expected to minimize the costs while providing satisfactory charging services and maintaining system reliability. More specifically, we propose an MARL-based approach to equip each EV charger with an energy management policy that can handle uncertainties from time-varying energy resources and charging demand, as well as system faults. Furthermore, we design a customized dense reward for the MARL-based approach to improve charging satisfaction. Long Short-Term Memory (LSTM) \cite{hochreiter1997long} is adopted as the main structure of the neural networks used in the MARL algorithm.
The main contributions of this paper are as follows.
\begin{itemize}
  \item \textit{Decentralized energy management for improved system reliability}: Our proposed MARL employs the paradigm of Centralized Training with Decentralized Execution (CTDE), aiming to eliminate the need for communication between agents (i.e., EV chargers) when making charging/discharging decisions. This enables each agent to operate independently, such that faulty chargers do not affect other functioning chargers, while still being trained to coordinate their decisions through interactive learning.
  \item \textit{LSTM for enhanced temporal feature extraction}: The proposed MARL algorithm harnesses the advantage of LSTM in capturing temporal patterns and long-term autocorrelations to process time series data for enhancing the MARL training and charging management performance.
  \item \textit{Dense reward design for enhanced user experience}: We introduce a dense reward signal to facilitate the completion of charging tasks, i.e., satisfying the charging demand. The results show that our MARL algorithm can increase the completion of charging tasks while minimizing energy costs.
\end{itemize}

\section{System Model}
\begin{figure}[b]
    \centering
    \includegraphics[width=0.9\linewidth]{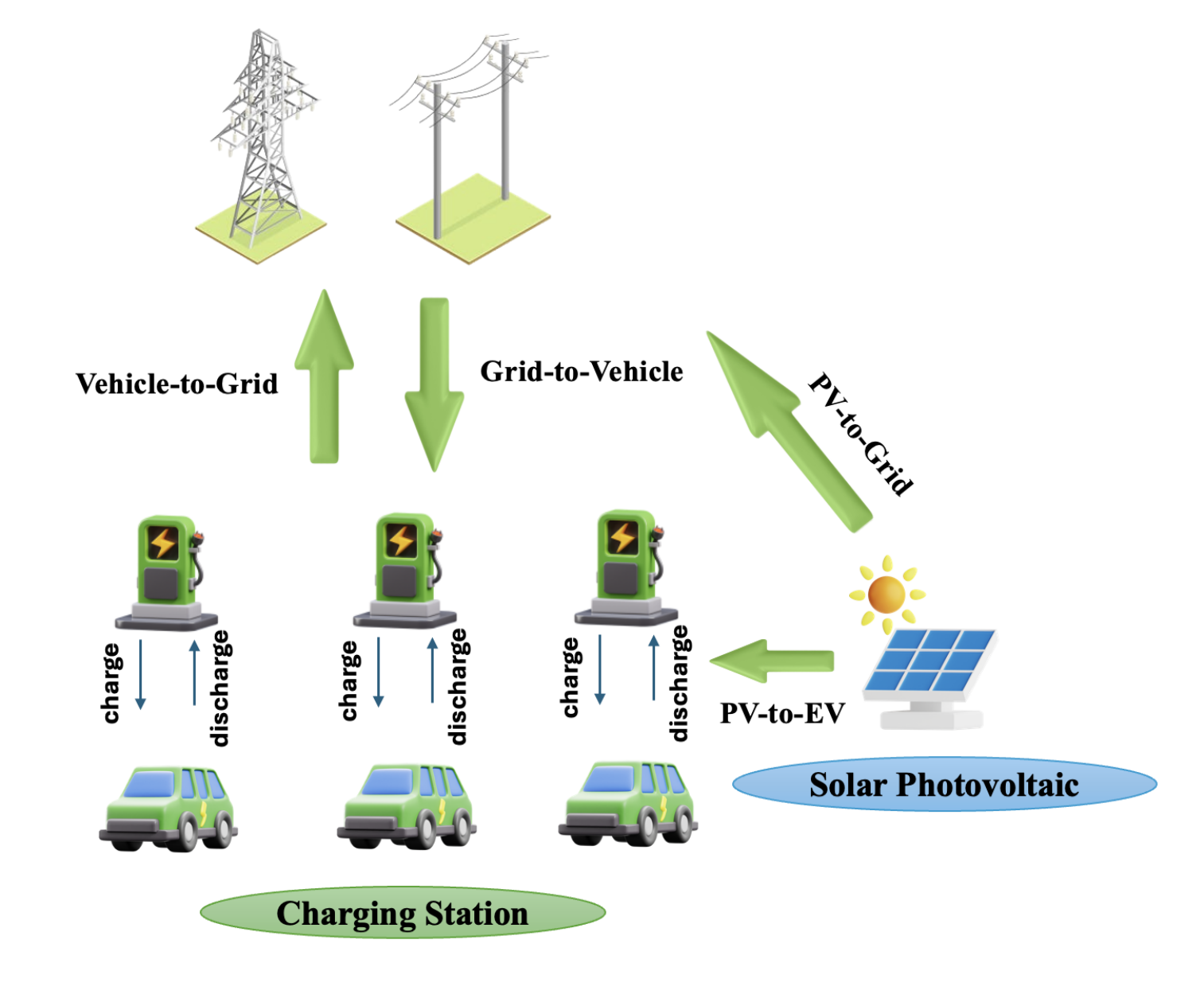}
    \caption{System model of an EV charging station with solar PV.}
    \label{sys}
\end{figure}
Fig. \ref{sys} shows a typical EV charging station with energy supply from onsite solar PV and the external grid. Within the EV charging station, there is a set of EV chargers $\mathcal{N} {=} \{1,..., N \}$, each of which can be connected to an EV. The operational horizon is denoted as $\mathcal{T}$ with discrete time slots $t \in \mathcal{T}$ with the same interval $\Delta t$.
The EV charging station aims to meet EV charging demands as much as possible, and any unmet charging demand leads to user dissatisfaction $ds_{i}$, which is formulated as
\begin{align}
    &  ds_{i}=\left\{
    \begin{array}{rcl}
    E_{i}^{\text{dem}} - E_{i,t=\text{tar}}    &      & E_{i,t=\text{tar}}     <      E_{i}^{\text{dem}}, \\
    0     &      & E_{i,t=\text{tar}}     \geq     E_{i}^{\text{dem}},\\
    \end{array} \right.
\end{align}
where $E_{i}^{\text{dem}}$ is the charging demand for EV $i$, and $E_{i,t=\text{tar}}$ is the energy level of EV $i$ at departure time $t=\text{tar}$. A larger unmet charging demand indicates greater dissatisfaction.

In addition to EV charging, V2G is also enabled in the charging station to discharge EV batteries to serve the grid when needed. We denote $a_{i,t}^\text{ch}$ and $a_{i,t}^\text{disch}$ as the charge and discharge of EV $i$ at $t$, and they satisfy the following constraints
\begin{align}
&0 \leq a_{i,t}^\text{ch} \leq \bar{P}^\text{ch} \label{xt_con},\\
    &0 \leq a_{i,t}^\text{disch} \leq \bar{P}^\text{disch} \label{xdt_con},
\end{align}
capped by the maximum charging power $\bar{P}^\text{ch}$ and discharging power $\bar{P}^\text{disch}$. The state-of-charge (SOC) for EV $i$ can be determined by $a_{i,t}^\text{ch}$ and $a_{i,t}^\text{disch}$ with charging and discharging efficiencies $\eta_i^\text{ch}$ and $\eta_i^\text{disch}$, but the SOC related constraints are omitted here due to limited space.

In the charging station, the power supply comes from the grid $a_{i,t}^\text{G2V}$ and PV $a_{i,t}^\text{PVEV}$ for EV $i$. Extra PV power in the charging station can feed into the grid, denoted by $a_{t}^\text{PVG}$, but the total PV power cannot exceed its generation $a_{t}^\text{PVgen}$. The total purchased power $\sum_{i \in \mathcal{N}}a_{i,t}^\text{G2V}$ or the sold power $\sum_{i \in \mathcal{N}} a_{i,t}^\text{disch} + a_{t}^\text{PVG}$ should not exceed the capacity of the charging station, denoted as $G^\text{max}$. All variables are non-negative. The above operational constraints are modeled as follows
\begin{align}
    & a_{i,t}^\text{ch} = a_{i,t}^\text{G2V} + a_{i,t}^\text{PVEV}, \label{ch_b} \\
    & 0 \leq  \sum_{i \in \mathcal{N}}a_{i,t}^\text{PVEV} + a_{t}^\text{PVG} \leq a_{t}^\text{PVgen}, \label{pvgen} \\
    & 0 \leq \sum_{i \in \mathcal{N}}a_{i,t}^\text{G2V} \leq G^\text{max},\label{g2v}\\
    & 0 \leq \sum_{i \in \mathcal{N}} a_{i,t}^\text{disch} + a_{t}^\text{PVG} \leq G^\text{max}, \label{v2g}\\
    & a_{i,t}^\text{PVEV}, a_{i,t}^\text{G2V}, a_{t}^\text{PVG} \geq 0. \label{non_neg}
\end{align}

V2G causes extra battery degradation, and it is important to model battery degradation costs when performing V2G. According to \cite{wu2022optimal}, the energy throughput equivalent method provides a suitable model. 
The cycle aging $AGE_{i,t}^\text{cyc}$ is modeled as
\begin{align}
    AGE_{i,t}^\text{cyc} = \frac{\lvert a_{i,t}^\text{ch} \eta^{\text{ch}}_i \Delta t -  \frac{a_{i,t}^\text{disch}\Delta t} {\eta^{\text{disch}}_i} \rvert}{2 \times E_i^\text{cap} L_i^\text{cyc}}, \label{AGE}
\end{align}
where the battery capacity is $E_i^\text{cap}$, and $L_i^\text{cyc}$ denotes the lifecycle parameter.

The objective of the charging station can be formulated as
\begin{align}
\begin{split}
&\sum_{t\in \mathcal{T}}\sum_{i \in \mathcal{N}} \Bigl( (a_{i,t}^\text{G2V} \kappa_t^\text{buy} - a_{i,t}^\text{disch}\kappa_t^\text{sell}) \Delta t + AGE_{i,t}^\text{cyc} \kappa^\text{batt}\Bigr) \\
&  + \sum_{i \in \mathcal{N}}ds_{i} - \sum_{t \in \mathcal{T}} (a_t^\text{PVG}\kappa_t^\text{sell} \Delta t),
\end{split}
\label{obj}
\end{align}
which includes net energy cost of EVs, denoted by $\sum_{t\in \mathcal{T}}\sum_{i \in \mathcal{N}} (a_{i,t}^\text{G2V} \kappa_t^\text{buy} - a_{i,t}^\text{disch}\kappa_t^\text{sell}) \Delta t$, battery degradation costs $\sum_{t\in \mathcal{T}}\sum_{i \in \mathcal{N}} AGE_{i,t}^\text{cyc} \kappa^\text{batt}$, user dissatisfaction $ \sum_{i \in \mathcal{N}} ds_{i}$, and solar power feed-in revenue $\sum_{t \in \mathcal{T}} (a_t^\text{PVG}\kappa_t^\text{sell} \Delta t)$. It is noted that the purchase price is $\kappa_t^\text{buy}$, the selling price is $\kappa_t^\text{sell}$, and the unit price of battery is $\kappa_t^\text{batt}$.

To implement our proposed MARL, we further formulate a multi-agent Markov decision process (MDP) for the above EV charging problem, including state, action and reward. We consider each EV charger as an agent. The input time-series data spans over a horizon $\mathcal{M}$. The state consists of the variables in the EV charging station as the environment, including the remaining charging demand of EVs $E_{i,t}^\text{r}$, remaining charging time $T_{i,t}^r$, energy purchase prices $\kappa_{t}^\text{buy}$, selling price $\kappa_{t}^\text{sell}$, and PV generation $a_t^\text{PVgen}$. 
For the action space, the charging station controls the charging/discharging of each connected EV, i.e., $a_{i,t} \in [-\bar{P}^\text{disch}, \bar{P}^\text{ch}]$.
The reward of agents usually aligns with the objective function in \eqref{obj}, but may require customized design. In particular, we include two parts in the reward function: cost function $R_{i,t}^\text{cost}$ and a customized charging satisfaction function $R_{i,t}^\text{user}$.
First, all agents aim to reduce energy costs under fluctuating PV generation and energy prices. The energy cost related reward is written as
\begin{align}
    & R_{i,t}^\text{cost} {=} -R_{i,t}^\text{cdb} + \frac{R_{i,t}^\text{pv}}{n},
\end{align}
which takes the negative form of costs, including the overall EV charging/discharging costs for each charger $i$ at time $t$, denoted as $R_{i,t}^\text{cdb}$, and the normalized PV feed-in revenue $R_{i,t}^\text{pv}$ divided by $n$-the number of chargers with EVs connected.

Regarding EV charging dissatisfaction, EV users only experience it at the time of departure and thus the reward can be sparse. It usually make it difficult for RL agent to learn effectively.
To address this challenge, we introduce a dense reward, designed to guide agents for completing the charging task. Specifically, the concept of completed average power is introduced as $CAP_{i,t} = \frac{E_{i,t}^\text{r}}{T_{i,t}^\text{r}}$, which measures the level of urgency to complete the charging task of EV $i$. An urgent status is identified when $CAP_{i,t}$ exceeds a threshold such as 80\% of the maximum charging, yet the current charging power $a_{i,t}$ remains below $CAP_{i,t}$. Then, a negative reward is sent to guide the agent to avoid this situation, expressed as
\begin{align}
    R_{i,t}^\text{user} {=} -(CAP_{i,t} - a_{i,t}).
\end{align}

\begin{figure*}[t]
    \centering
    \includegraphics[width=0.90\linewidth]{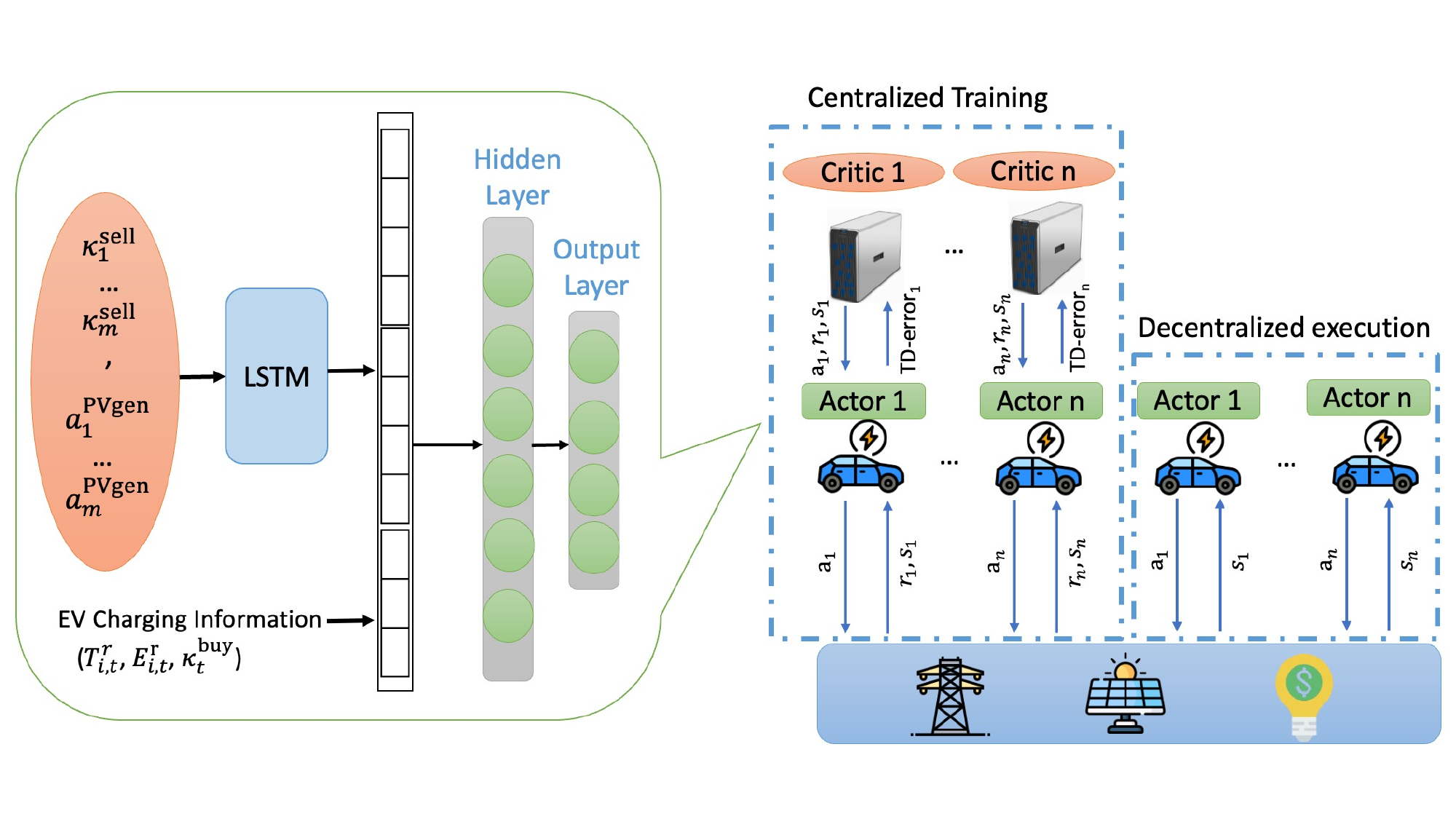}
    \caption{The neural network architecture (left) and the framework of MADDPG (right).}
    \label{meth}
\end{figure*}

Then, the final weighted reward $R_{i,t}$ is written as
\begin{align}
    & R_{i,t} = R_{i,t}^\text{cost} +  R_{i,t}^\text{user} - R^\text{grid},
\end{align}
where $R^\text{grid}$ is the penalty of grid constraint violation, if the purchased power or the sold power exceeds $G^\text{max}$.

\section{Methodology}
The proposed MARL algorithm employs centralized training with decentralized execution on the Actor-Critic structure using Multi-Agent Deep Deterministic Policy Gradient (MADDPG), which is shown on the right-hand side in Fig. \ref{meth}. LSTM-based neural networks are used, as shown on the left-hand side. For each charger, a centralized critic collects the information of all chargers and holistic system information from the grid to update the actor for maximizing rewards. For the actor, actions are executed independently to control the corresponding EV charger \cite{lowe2017multi}.

To better capture temporal features for each EV charger, LSTM is adopted as the primary structure of the neural networks to process PV generation and electricity price information, followed by a fully connected layer to output actions.
The critic and the actor of each EV charger have their own neural networks. The critic accesses all chargers' information and uses the neural network to approximate the Q-function, which evaluates the quality of actions. For each actor, its neural network is used to learn its policy, while only the local information (of each charger itself) is accessible to the actor.

The agent can learn policies that output continuous actions, e.g., EV charging and discharging. For each continuous policy $\mu_i$, the loss function is shown below
\begin{equation}
\begin{aligned}
\nabla_{\theta_i} J\left(\theta_i\right)=& \mathbb{E}_{\textbf{s}, a_i \sim D}\left[\nabla_{\theta_i} \mu_i\left(a_i \mid s_i\right)\right.\\
&\left.\left.\nabla_{a_i} Q_i^\mu\left(\textbf{s}, \textbf{a}\right)\right|_{a_i=\mu_i(s_i)}\right].
\end{aligned}
\end{equation}
The replay buffer $D$ includes state, action, and reward information. The state and action of all agents are represented as $\textbf{s}$ and $\textbf{a}$, respectively. The centralized Q-function of deterministic policy is denoted as $Q_i^{\mu}(\textbf{s},\textbf{a})$.

For centralized training of critics, the parameters in $i$'s critic network can be updated by Temporal Difference (TD) learning. The output of the Q-function is the score of the corresponding action. 
The update rules of the loss function are provided in \cite{lowe2017multi}.
For training, each critic receives all states in the grid and actions of all chargers to guide its own actor to learn collaborative policies. Then, each charger uses its policy to output charging power based on its individual state and action.
After training, the policy network associated with each charger is equipped to render what is approximately the optimal decision concerning the output of charging power. 
When these policy networks are implemented across distributed charging stations, they endow each agent with the ability to make autonomous decisions pertaining to its respective charger, thus improving the robustness of the holistic grid with multiple EV chargers when facing uncertain partial faults.

\begin{figure}[t]
    \centering
    \includegraphics[width=7.0cm]{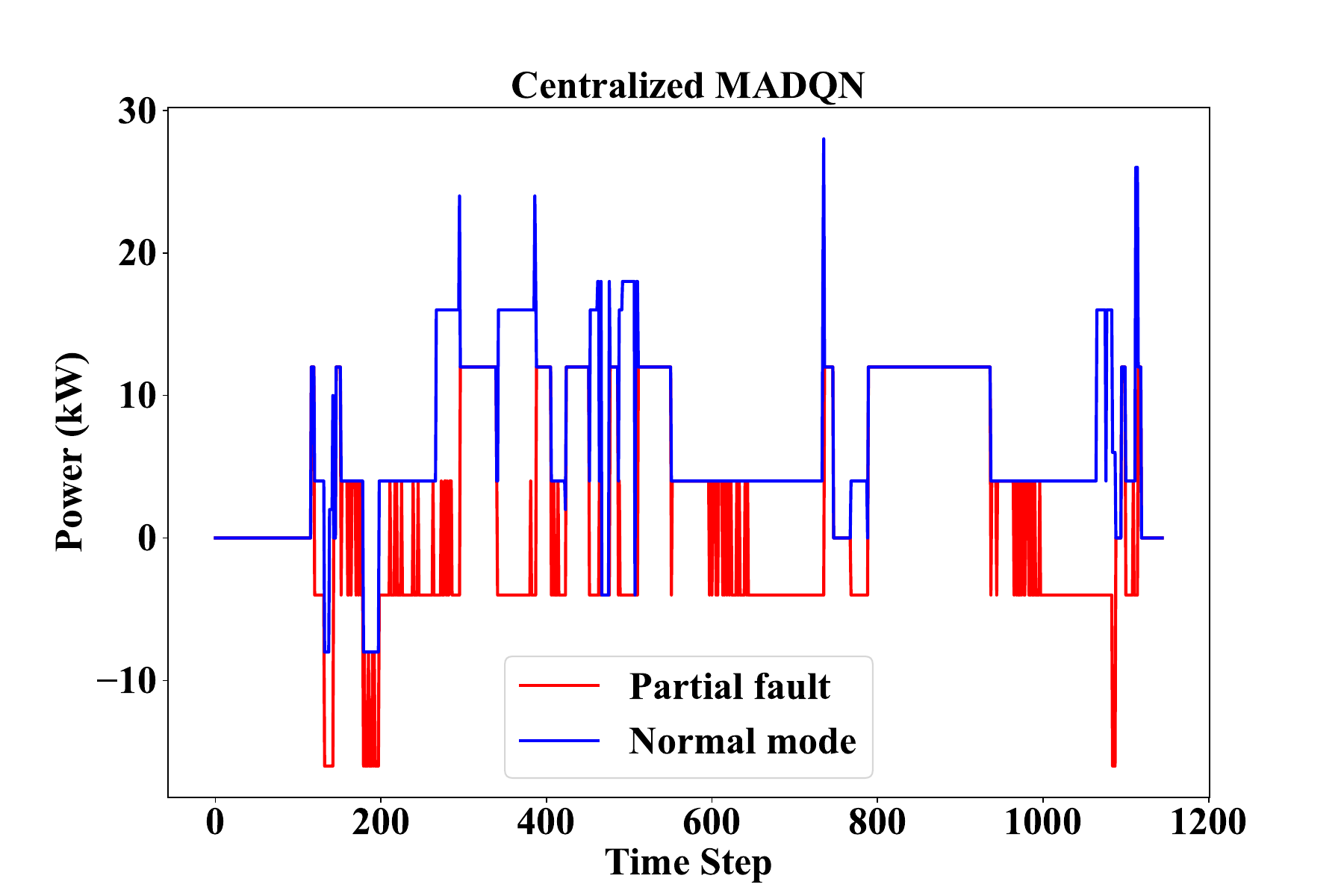}
    \caption{The agent's action for centralized MARL in the case of partial faults.}
    \label{re}
\end{figure}
\begin{figure}[t]
    \centering
    \includegraphics[width=7.0cm]{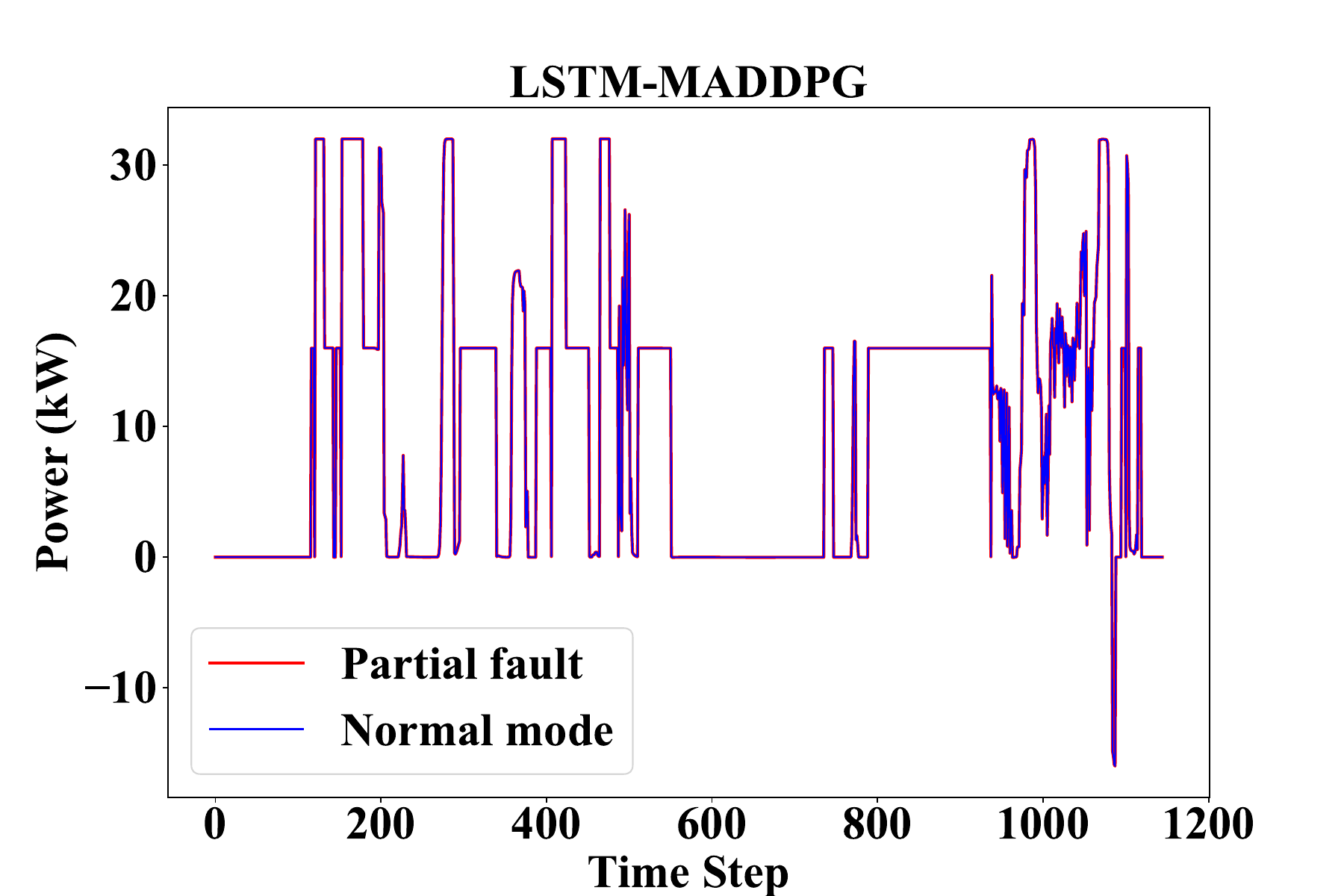}
    \caption{The agent's action for decentralized MARL in the case of partial faults.}
    \label{re1}
\end{figure}

\begin{figure}[t]
    \centering
    \includegraphics[width=7.0cm]{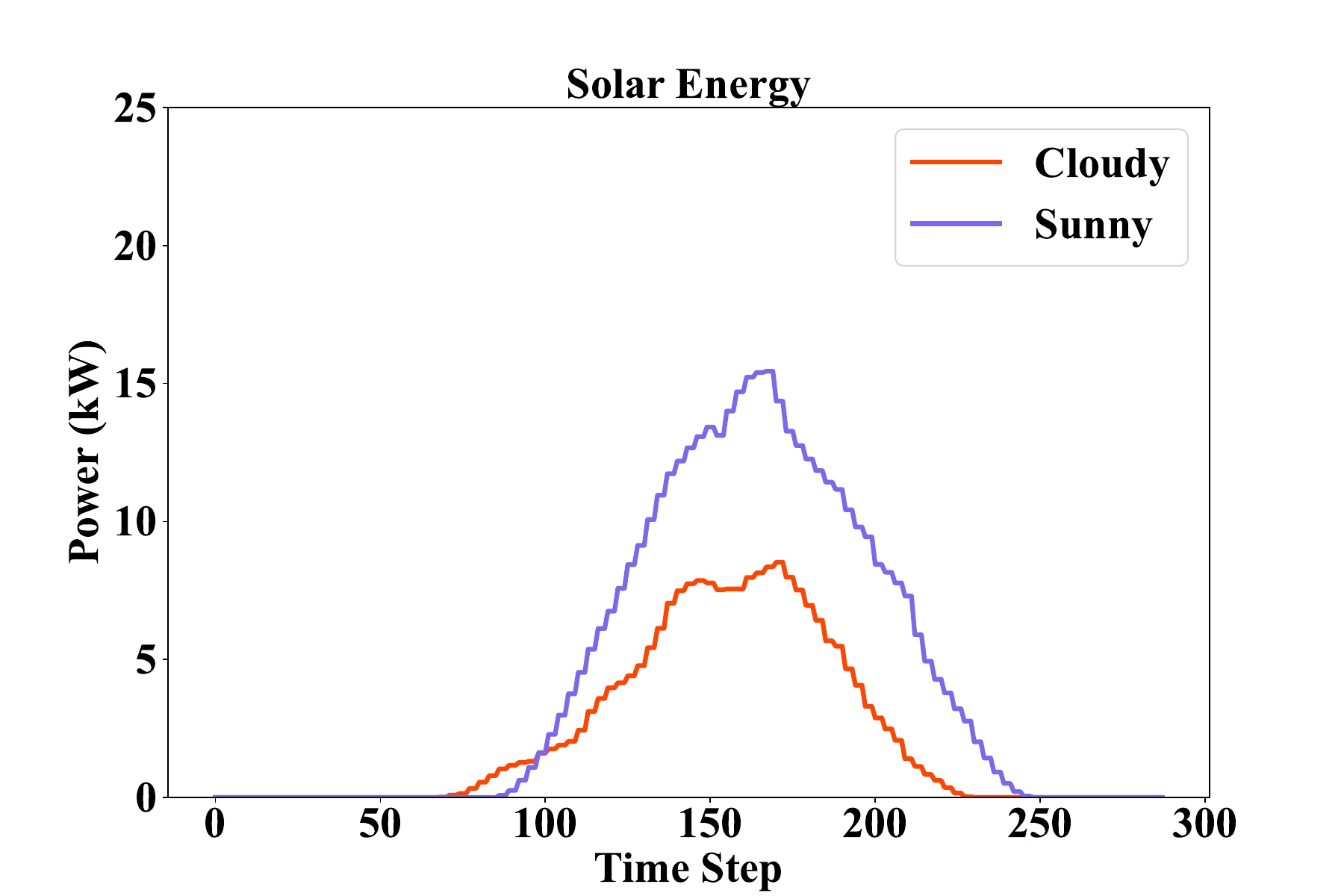}
    \caption{Solar energy under different weather conditions.}
    \label{solar}
\end{figure}

\begin{table}
\centering
\caption{Performance for Different Algorithms.}
\begin{tabular}{ |c|c|c| } 
 \hline
 Algorithm & Energy Cost & Unfinished Demand \\ 
 \hline
 MADQN & 1361.6 & 82.1 \\ 
 MADDPG &1102.3&584.5\\
 LSTM-MADDPG (Sparse reward)& 1086.2 & 349.6 \\
 \textbf{LSTM-MADDPG (Dense reward)} & 1021.6 & 28.9 \\
 \hline
\end{tabular}
\label{result}
\end{table}

\section{Simulations and Results}
We set up a charging station with $20$ chargers, two of which are unreliable. The unreliable chargers may be attacked to generate random data to disrupt the system. We use EV data in Los Angeles from September 2018 to June 2019 \cite{lee2019acn}. Solar data \cite{Elia_2021} is used to simulate our solar power system. We set the solar capacity to $32$kWh and the maximum charging/discharging power is $22$kW.
Time-Of-Use (TOU) price is used for the purchase price and the wholesale energy price \cite{aemo_dashboard} is used for selling energy. 

We implement different MARL algorithms as the baseline, including centralized multi-agent deep Q-network (MADQN),  MADDPG and LSTM-MADDPG with CTDE under different reward settings. We evaluate the performance concerning unfinished EV charging demand and the total energy cost in one month. The battery degradation is included in the energy cost. Lower energy costs and unfinished demand indicate better performance.

Table \ref{result} demonstrates the performance between different algorithms. We can see that our proposed LSTM-MADDPG performs better than other baseline algorithms in terms of the energy cost and unfinished demand. Moreover, we compare the performance of LSTM-MADDPG under different reward design, our dense reward design can reduce the unfinished demand significantly. At last, since LSTM is used to capture time series information, our proposed LSTM-MADDPG achieves the best performance in terms of both energy cost and unfinished demand than MADDPG.

\begin{figure}[t]
    \centering
    \includegraphics[width=7.0cm]{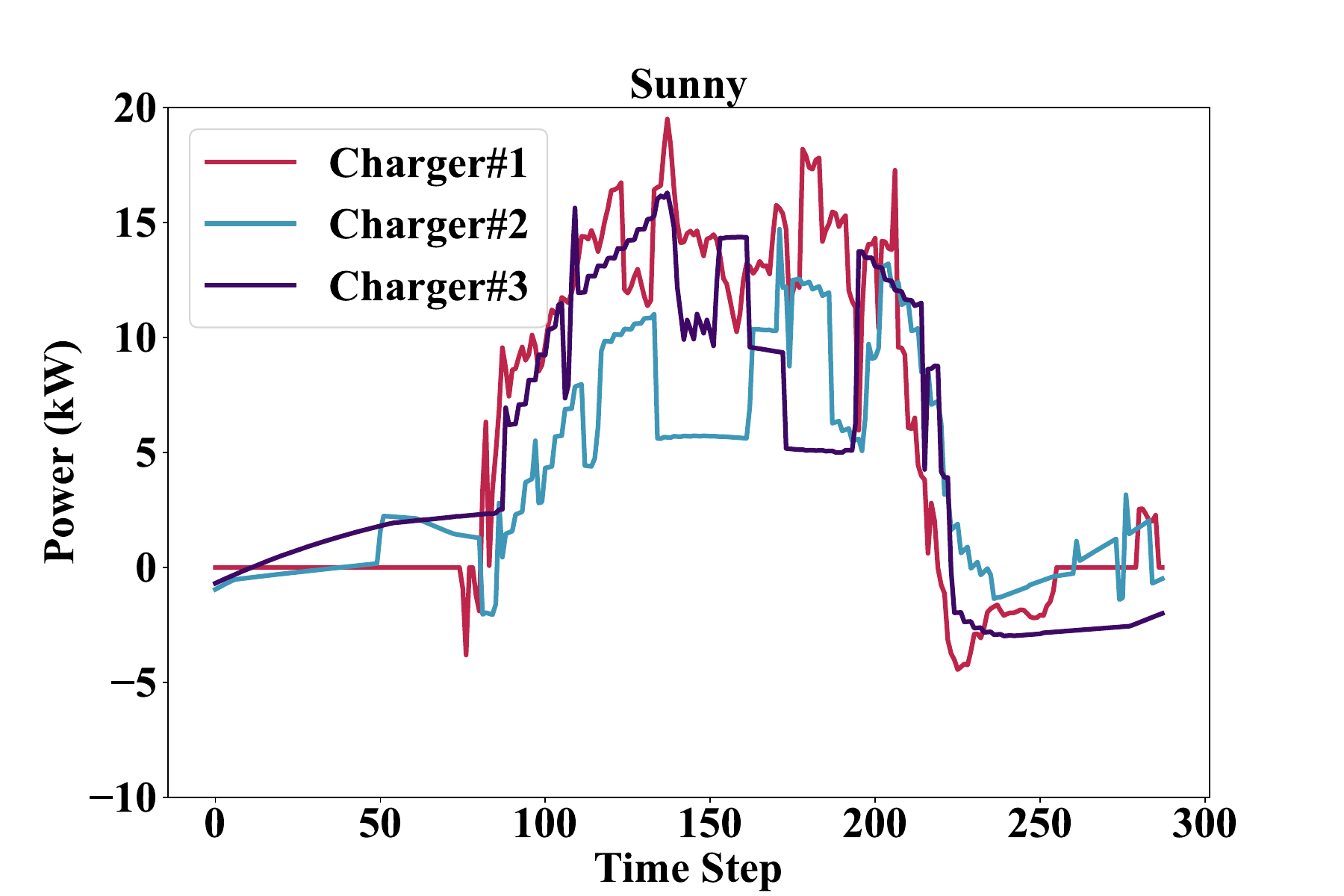}
    \caption{Three chargers' decisions under sunny conditions using LSTM-MADDPG.}
    \label{time1}
\end{figure}
\begin{figure}[t]
    \centering
    \includegraphics[width=7.0cm]{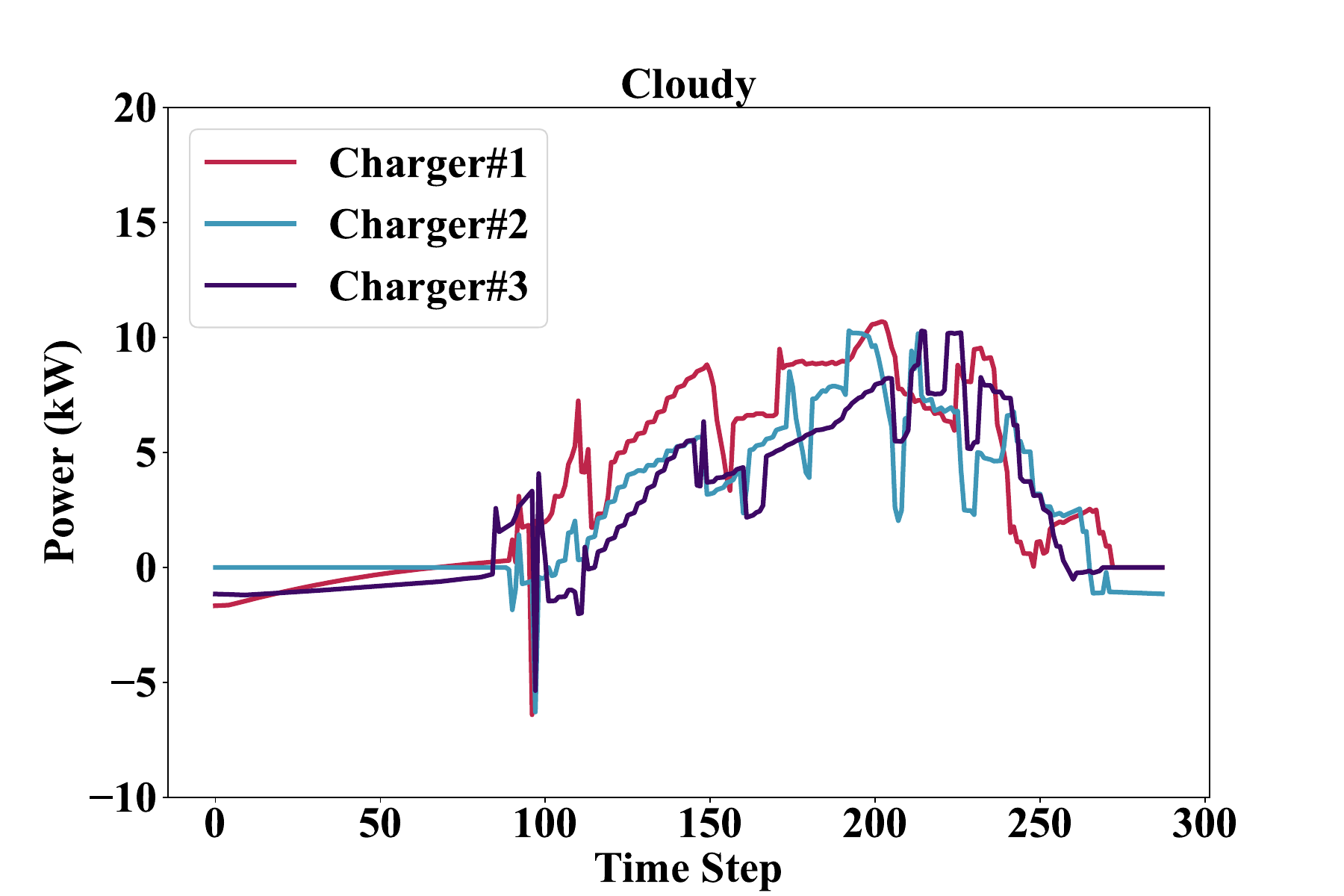}
    \caption{}
    \label{time2}
\end{figure}

The reliability of CTDE in our LSTM-MADDPG is demonstrated in Fig. \ref{re} and Fig. \ref{re1}. We simulate partial charger faults using randomly generated values to replace the information of faulty chargers. The normal mode without charger faults is used to simulate chargers working normally. The red line and blue line show the actions of working chargers, such as charge and discharge, in faulty and normal modes. Fig. \ref{re} shows that the partial fault causes the instability of centralized MADQN, while our proposed LSTM-MADDPG with CTDE is working as normal, as depicted in Fig. \ref{re1}.

We also show how our LSTM-MADDPG algorithm performs under different weather conditions, e.g., on cloudy and sunny days, as shown in Fig. \ref{solar}, in which the cloudy weather will reduce solar energy generation. Fig. \ref{time1} and Fig. \ref{time2} show the multi-agent decisions under different weather conditions. We can see that the proposed algorithm can learn the pattern of solar energy generation under different weather conditions and charge more on the sunny day to maximize solar energy utilization.

\section{Conclusion}
Multi-agent reinforcement learning with LSTM and CTDE is proposed for EV charging coordination in a charging station. The decentralized management strategy ensures the reliability of the charging service even in the event of system faults. This paper uses LSTM to encode time series data, enabling the agent to learn the pattern of time series and improve the algorithm's performance. This work also improves charging satisfaction by designing a dense reward function to deal with sparse signals of charging satisfaction.
For future research, we will consider power network constraints, including voltage constraints. Future research will also aim to solve a larger-scale problem for charging station energy management in a more realistic setting.

% References
\bibliographystyle{IEEEtran}
\bibliography{IEEEabrv.bib,ref.bib}

\end{document}